# Topological phase transitions in a honeycomb ferromagnet with unequal Dzyaloshinskii-Moriya interactions

Heng Zhu[a], Hongchao Shi[a], Zhengguo Tang, Bing Tang[*]

*Department of Physics, Jishou University, Jishou 416000, China*

ABSTRACT

This theoretical research is devoted to study topological phase transitions in a two-dimensional honeycomb ferromagnetic lattice with unequal Dzyaloshinskii-Moriya interactions for the two sublattices. With the help of a first-order Green's function formalism, we analyze the influence of magnon-magnon interaction on the magnon band topology. It is found that the existence of the antichiral Dzyaloshinskii-Moriya interaction can led to a tilting of the renormalized magnon bands near the Dirac momenta. Then, the renormalized magnon band gaps at Dirac points have different widths. Through changing the temperature, we can observe the renormalized magnon band gap closing-reopening phenomenon, which corresponds to the topological phase transition. Our results show that the critical temperature of the topological phase transition is related to the strength of the antichiral Dzyaloshinskii-Moriya interaction.

## 1. Introduction

During the past decade, the quantum ferromagnet or antiferromagnet has appeared as a multifunction stage to realize the magnetic analog of the topological phase in electronic systems [1-14]. Since topological properties of two-dimensional (2D) electronic systems (i.e., graphene) have been extensively investigated, more and more attention has been paid to their boson analogues in various platforms, such as photonic [15-17], phononic [18,19], and magnonic systems [20-23]. Physically, magnons are the quantum counterparts of (linear) spin waves, which are the collective excitation modes in magnets [24,25]. Different from electrons, magnons have no electrical charge so that forming one magnon current does not invite Joule heating, which has application potential in the realization of the low-dissipation device [21,26-28].

[*] Corresponding author.
E-mail addresses: bingtangphy@jsu.edu.cn
[a] These authors contributed equally to this work.

In fact, it is easy to control magnetic properties via applied magnetic fields, which means that the magnonic band provides one unique platform to probe the plentiful and still developing basic principles of the band theory. In ferromagnets or antiferromagnets, the presence of the Dzyaloshinskii-Moriya interaction (DMI) shall destroy the spatial inversion symmetry of the system, which can give rise to the nontrivial topological magnon band structure and corresponding nonzero Berry curvature [3-5,28-30]. As (uncharged) magnons are not subjected to the Lorentz force, the DMI shall play the role of the effective magnetic field in momentum space via affecting the magnon motion in the magnetic system, which can cause a thermal magnon Hall effect [28,29]. The thermal magnon Hall effect caused by the DMI has been first predicted theoretically in the kagome and pyrochlore ferromagnets [3]. Afterwards, Onose *et al.* [4] have experimentally observed the thermal magnon Hall effect in one ferromagnetic insulator $Lu_2V_2O_7$ with three-dimensional pyrochlore structure. Subsequently, the thermal magnon Hall effect has been experimentally identified in one two-dimensional kagome magnet Cu(1-3, bdc) [31]. In addition, the thermal magnon Hall effect in honeycomb magnets has also been theoretically predicted [29]. In honeycomb ferromagnet, one magnon band gap opens at six Dirac points as the inversion symmetry of the system has been broken by the second nearest neighbor DMI, which is analogous to the effect of the spin-orbit interaction in the Kane-Mele model on one honeycomb lattice [2].

In contrast to those electronic systems, the deficiency of the magnon number conservation allows for the nonconserving many-magnon interactions and spontaneous decays [32,33]. Unfortunately, a majority of theoretical works on the magnonic band topology have been only focused on the linear spin wave approximation, in which the magnon-magnon interaction has been ignored [8,29,34]. When the temperature is very low, there is not much debate about that the magnon-magnon interaction effect is thought to be frozen out [35]. With the increase of temperature, the role of the magnon-magnon interaction has to be considered. What is wore, for the sake of evaluating the Hamiltonian including magnon-magnon interactions, those systematic perturbative expansions must be taken into account. Recently, the significance of the many-body effect has been recognized in magnonic Dirac systems [32,35-37]. Pershoguba *et al.* [32] have first studied the effect of the magnon-magnon interactions on Dirac magnons in two-dimensional honeycomb ferromagnets and showed that such interactions can give rise to the remarkable momentum-dependent renormalization of band structures. Their theory has perfectly worked out abnormal phenomena in the neutron-scattering experiment for $CrBr_3$, which has not been explained for nearly half a century. It should be mentioned that the DMI has been ignored in their theory, though it has been experimentally identified in $CrI_3$ and $CrBr_3$ [38,39]. Fortunately, Mook *et al.* [35] have showed that the orientation of the DMI can affect the magnetic topological characteristics and forms of the magnon-magnon interactions in the honeycomb ferromagnet. Moreover, Lu *et al.* [36] have incorporated a chiral DMI and the magnon-magnon interaction into the system Hamiltonian in order to study topological physics of the honeycomb ferromagnets. Their results have indicated that the magnon-magnon interactions can cause topological phase transitions in the honeycomb ferromagnet, which are portrayed via the sign variation of the magnon thermal Hall conductivity. Very recently, Sun *et al.* have studied the influence of interactions on topological magnons in a honeycomb ferromagnet with DMI [37]. By making use of a parametric magnon amplification scheme, the tunable renormalization has been proposed. They have showed that the magnonic Dirac material can be viewed as a significant platform to explore many-body physics of bosonic systems having nontrivial topological band structures.

In this work, we study topological magnons in one Heisenberg honeycomb ferromagnet with unequal DMIs on the different sublattices. Inspired by the ideas form Refs. [36,37], the effect of magnon-magnon interaction on the band topology shall be analyze by making use of Green's function formalism up to the first-order in the perturbation theory. The existence of the antichiral DMI destroys the chiral symmetry of the honeycomb ferromagnet, which can cause a tilting of the renormalized magnon bands near the Dirac momenta. What is more, the renormalized magnon band gaps at non-equivalent Dirac points are no longer equal with the increase of the temperature. By varying the temperature, we can observe the band gap closing-reopening phenomenon, which is signature for the topological phase transition. The role of the antichiral DMI in the topological phase transition of Dirac magnons will be discussed. More details on this work will be fully displayed in the following sections.

**2. Model**

Let us consider a two-dimensional Heisenberg honeycomb ferromagnet with unequal second nearest-neighbor DMIs on the A and B sublattices, as shown in Fig. 1. The symmetric and antisymmetric combinations of $D_A$ and $D_B$ are defined as $D = \frac{1}{2}(D_A + D_B)$ and $D' = \frac{1}{2}(D_A - D_B)$, which are called chiral and antichiral DMI, respectively [40]. In the presence of an external magnetic field, the complete spin Hamiltonian of such a system can be written in the following form

$$\mathcal{H} = -J\sum_{\langle i,j \rangle} \vec{S}_i \cdot \vec{S}_j + D\sum_{\langle\langle i,j \rangle\rangle} v_{ij}\hat{z} \cdot \vec{S}_i \times \vec{S}_j + D'\sum_{\langle\langle i,j \rangle\rangle} v'_{ij}\hat{z} \cdot \vec{S}_i \times \vec{S}_j - B\sum_i S_i^z . \quad (1)$$

where the first term corresponds to the nearest-neighbor ferromagnetic exchange interaction, $J > 0$. The second and third terms represent the chiral DMI and antichiral DMI between two next-nearest spins, respectively. Here, the DM vectors are constrained to the Z-axis positive direction, where the constants $v_{ij} = -v_{ji} = \pm 1$ depend on the relative position of two next-nearest sites and $v'_{ij} = +v_{ij}$ for sublattice A and $v'_{ij} = -v_{ij}$ for sublattice B. The last term is caused by the Zeeman effect.

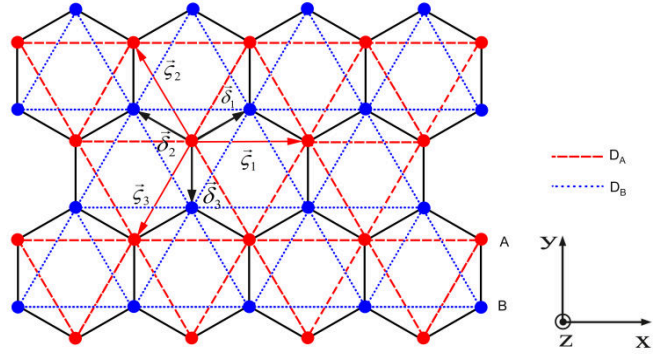

**Fig. 1.** Schematic of the honeycomb lattice structure, which consists of two triangular sublattices. $\vec{\delta}_n$ and $\vec{\varsigma}_n$ ($n=1,2,3$) denote the three nearest-neighbor and next-nearest-neighbor vectors, respectively.

In order to bosonize the model spin Hamiltonian, we shall adopt the Holstein-Primakoff (HP) transformation [41]. In the low temperature limit, $2S \gg \langle \hat{n} \rangle = \langle a_i^+ a_i \rangle$, and the square roots can be expanded in powers of $1/\sqrt{S}$. Truncated to the order of $S^{\frac{1}{2}}$, these transformations can be simplified as $S_i^+ = \sqrt{2S}a_i$, $S_i^- = \sqrt{2S}a_i^+$, and $S_i^z = S - a_i^+ a_i$. Thus, one can get the noninteracting bosonic Hamiltonian $\mathcal{H}_0 = \sum_{\vec{k}} \psi_{\vec{k}}^+ H_0(\vec{k}) \psi_{\vec{k}}$, where $\psi_{\vec{k}}^+ = (a_{\vec{k}}^+, b_{\vec{k}}^+)$ is a spinor denoting the degrees of freedom for the two sublattices, and $H_0(\vec{k})$ reads

$$H_0(\vec{k}) = h_0(\vec{k})\sigma_0 + h_x(\vec{k})\sigma_x + h_y(\vec{k})\sigma_y + h_z(\vec{k})\sigma_z. \qquad (2)$$

Here, $h_0(\vec{k}) = 3JS + B + 2D'S\rho_{\vec{k}}$, $h_x(\vec{k}) = -JS\,\mathrm{Re}(\gamma_{\vec{k}})$, $h_y(\vec{k}) = JS\,\mathrm{Im}(\gamma_{\vec{k}})$, $h_z(\vec{k}) = 2DS\rho_{\vec{k}}$, $\gamma_{\vec{k}} = \sum_{n=1}^{3} e^{i\vec{k}\cdot\vec{\delta}_n}$, $\rho_{\vec{k}} = \sum_{n=1}^{3} \sin(\vec{k}\cdot\vec{\varsigma}_n)$, and the vectors $\vec{\delta}_n$ and $\vec{\varsigma}_n$ are shown in Fig.1. $\sigma_0$ and $\sigma_\alpha (\alpha = x, y, z)$ correspond to the unit and Pauli matrices, respectively. By diagonalizing $H_0(\vec{k})$, one can obtain the magnon dispersion relation, which is

$$\varepsilon_\alpha^0(\vec{k}) = h_0(\vec{k}) + \eta\varepsilon(\vec{k}), \qquad (3)$$

where $\varepsilon(\vec{k}) = \sqrt{h_x(\vec{k})^2 + h_x(\vec{k})^2 + h_z(\vec{k})^2}$ and $\eta = 1(-1)$ corresponds to the up(down) band, namely, $\alpha = u(\alpha = d)$.

In Fig. 2, we display the magnon (linear) dispersion relation of the present honeycomb ferromagnet. In the inexistence of DMI, the two bands meet at two inequivalent Dirac points, namely, $\vec{K}_+ = \left(\frac{4\sqrt{3}\pi}{9a}, 0\right)$ and $\vec{K}_- = \left(\frac{2\sqrt{3}\pi}{9a}, \frac{2\pi}{3a}\right)$. A nonzero $D$ can cause an effective Haldane mass term that opens up a non-trivial band gap $\Delta = 6\sqrt{3}DS$ between the upper and lower branches at these Dirac points.

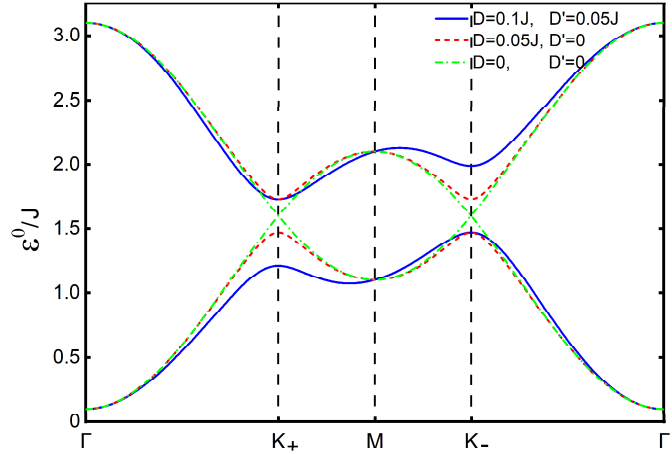

**Fig. 2.** The magnon linear dispersion relation along the path $\Gamma - K_+ - M - K_- - \Gamma$. The magnetic induction intensity is set to $B = 0.1J$ and $S = \frac{1}{2}$.

When there exists unequal DMIs on the A and B sublattices ($D_A \neq D_B$), namely, the existence of the antichiral DMI, the magnon band gap opening is not symmetric and leads to a tilting of the magnon bands around these Dirac points, as displayed in Fig. 2. Thus, we conclude that the antichiral DMI can cause the tilting of the magnon bands. However, the antichiral DMI has no influence on the width of the magnon band gap.

**3. Interacting topological Dirac magnons**

*3.1. The renormalization of magnon band*

Utilizing the HP transformation to expand the original Hamiltonian to the order of $\frac{1}{\sqrt{S}}$, one can get the two-magnon interacting Hamiltonian in the original ab basis, which is given by

$$\mathcal{H}_{int} = \frac{J}{4N}\sum_{\{\vec{k}_i\}}\left(\gamma^*_{\vec{k}_4} b^+_{\vec{k}_1} b^+_{\vec{k}_2} b_{\vec{k}_3} a_{\vec{k}_4} + \gamma_{\vec{k}_2} b^+_{\vec{k}_1} a^+_{\vec{k}_2} b_{\vec{k}_3} b_{\vec{k}_4} + \gamma_{\vec{k}_4} a^+_{\vec{k}_1} a^+_{\vec{k}_2} a_{\vec{k}_3} b_{\vec{k}_4} + \gamma^*_{\vec{k}_2} a^+_{\vec{k}_1} b^+_{\vec{k}_2} a_{\vec{k}_3} a_{\vec{k}_4} - 4\gamma_{\vec{k}_4-\vec{k}_2} a^+_{\vec{k}_1} b^+_{\vec{k}_2} a_{\vec{k}_3} b_{\vec{k}_4}\right)$$
$$- \frac{D+D'}{2N}\sum_{\{\vec{k}_i\}}\left(\rho_{\vec{k}_2} + \rho_{\vec{k}_4}\right) a^+_{\vec{k}_1} a^+_{\vec{k}_2} a_{\vec{k}_3} a_{\vec{k}_4} + \frac{D-D'}{2N}\sum_{\{\vec{k}_i\}}\left(\rho_{\vec{k}_2} + \rho_{\vec{k}_4}\right) b^+_{\vec{k}_1} b^+_{\vec{k}_2} b_{\vec{k}_3} b_{\vec{k}_4},$$

(4)

where $\{\vec{k}_i\}$ stands for summation over all $\vec{k}_i$. In Eq. (4), the conservation of momentum is due to $\frac{1}{N}\sum_i e^{i(\vec{k}_1+\vec{k}_2-\vec{k}_3-\vec{k}_4)\cdot\vec{r}_i} = \delta_{\vec{k}_1+\vec{k}_2,\vec{k}_3+\vec{k}_4}$. These interaction terms can be rewritten as a more compact expression, namely, $\mathcal{H}_{int} = \sum_{\{\vec{k}_i\}} V^{\vec{k}_1,\vec{k}_2}_{\vec{k}_3,\vec{k}_4} \psi^+_{\vec{k}_1}\psi^+_{\vec{k}_2}\psi_{\vec{k}_3}\psi_{\vec{k}_4}$. Here, the summation runs over all wavevectors $(\vec{k}_1,\vec{k}_2,\vec{k}_3,\vec{k}_4)$ compositions, which are constrained via the momentum conservation.

According to the perturbation technique for the single-magnon Green's function developed by Sun *et al.*[37], one can obtain a first order Dyson equation on interacting magnons, which has the following form

$$G_R(\vec{k},\vec{k}';\omega) = G^{(0)}_R(\vec{k},\vec{k}';\omega) + G^{(0)}_R(\vec{k},\vec{k}';\omega)\sum\nolimits^{(1)}_{H_{int}}(\vec{k})G^{(0)}_R(\vec{k},\vec{k}';\omega). \tag{5}$$

Here, $G_R(\vec{k},\vec{k}';\omega)$ is called the interacting Green's function,

$G_R^{(0)}(\vec{k},\vec{k}';\omega) = \dfrac{\delta_{\vec{k},\vec{k}'}}{\omega - H_0(\vec{k})}$ is known as the free magnon Green's function, and $\sum_{H_{int}}^{(1)}(\vec{k})$ stands for the first-order self-energy.

By expanding $[\psi_{\vec{k}}, \mathcal{H}_{int}]$, one can get the Hartree-type self-energy, namely,

$$\sum\nolimits_{H_{int}}^{(1)}(\vec{k}) = \begin{pmatrix} Q_{\vec{k}} - \delta_m & M\gamma_{\vec{k}} - p_{\vec{k}} \\ M\gamma_{\vec{k}}^* - p_{\vec{k}}^* & Q_{\vec{k}} + \delta_m \end{pmatrix} \qquad (6)$$

with

$$M = \frac{J}{2N}\sum_{\vec{q}}\mu^+(\vec{q}) \quad , \quad p_{\vec{k}} = \frac{J}{2N}\sum_{\vec{q}}\gamma_{\vec{k}-\vec{q}}e^{i\phi_{\vec{q}}}\sqrt{1-\chi(\vec{q})^2}\mu^-(\vec{q}),$$

$$Q_{\vec{k}} = \frac{J}{2N}\sum_{\vec{q}}\left(|\gamma_{\vec{q}}|\sqrt{1-\chi(\vec{q})^2}\mu^-(\vec{q}) - \gamma_0\mu^+(\vec{q})\right) + \frac{D}{N}\sum_{\vec{q}}\rho_{\vec{q}}\chi(\vec{q})\mu^-(\vec{q}) - \frac{D'}{N}\sum_{\vec{q}}\rho_{\vec{q}}\mu^+(\vec{q})$$

$$+\frac{D}{N}\rho_{\vec{k}}\sum_{\vec{q}}\chi(\vec{q})\mu^-(\vec{q}) - \frac{D'}{N}\rho_{\vec{k}}\sum_{\vec{q}}\mu^+(\vec{q}),$$

$$\delta_m = \frac{J\gamma_0}{2N}\sum_{\vec{q}}\chi(\vec{q})\mu^-(\vec{q}) + \frac{D}{N}\sum_{\vec{q}}\rho_{\vec{q}}\mu^+(\vec{q}) - \frac{D'}{N}\sum_{\vec{q}}\rho_{\vec{q}}\chi(\vec{q})\mu^-(\vec{q})$$

$$+\frac{D}{N}\rho_{\vec{k}}\sum_{\vec{q}}\mu^+(\vec{q}) - \frac{D'}{N}\rho_{\vec{k}}\sum_{\vec{q}}\chi(\vec{q})\mu^-(\vec{q}),$$

$$\mu^+(\vec{q}) = f(\varepsilon_d^0(\vec{q})) + f(\varepsilon_u^0(\vec{q})) \quad , \quad \mu^-(\vec{q}) = f(\varepsilon_d^0(\vec{q})) + f(\varepsilon_u^0(\vec{q})). \qquad (7)$$

Here, $f(\varepsilon_\alpha^0(\vec{q})) = \left[e^{\beta\varepsilon_\alpha^0(\vec{q})} - 1\right]^{-1}$ is the Bose-Einstein distribution function, $\beta = 1/k_B T$, and $\chi(\vec{q}) = \dfrac{h_z(\vec{q})}{\varepsilon(\vec{q})}$.

In the low temperature, the second-order effect for the DMI can be neglected [37]. Here, our aim is to explore the effect of the antichiral DMI on the topological property of the present model. Hence, following Ref. [36], we will only consider the first-order renormalized Hamiltonian. It is not difficult to obtain the renormalized Hamiltonian for our system, which is

$$H_1(\vec{k}) = H_0(\vec{k}) + \sum\nolimits_{H_{int}}^{(1)}(\vec{k})$$
$$= \begin{pmatrix} 3JS + B + 2D'S\rho_{\vec{k}} + Q_{\vec{k}} + m_{\vec{k}} & -(JS - M)\gamma_{\vec{k}} - p_{\vec{k}} \\ -(JS - M)\gamma_{\vec{k}}^* - p_{\vec{k}}^* & 3JS + B + 2D'S\rho_{\vec{k}} + Q_{\vec{k}} - m_{\vec{k}} \end{pmatrix}.$$

(8)

Here, $m_{\vec{k}} = 2DS\rho_{\vec{k}} - \delta_m$ is the total mass term.

By diagonalizing the renormalized Hamiltonian (8), one can obtain a renormalized magnon dispersion relation, which reads

$$\varepsilon_\alpha(\vec{k}) = 3JS + B + 2D'S\rho_{\vec{k}} + Q_{\vec{k}} \pm \sqrt{\left|(JS - M)\gamma_{\vec{k}} + p_{\vec{k}}\right|^2 + m_{\vec{k}}^2}. \quad (9)$$

Physically, the topological nature of Dirac magnons is associated with the interaction-induced Haldane mass term. It is easy to conclude that $\delta_m$ is numerically correlated with $D$, namely, $\delta_m \propto D$. Under the limitation $N \to \infty$, we have $\frac{1}{N}\sum_{\vec{q}} = \frac{A}{(2\pi)^2}\int_{BZ} d^2\vec{q}$. Here, $A = \frac{3\sqrt{3}a^2}{2}$ stands for the area of the unit cell of honeycomb lattice. Here, the summation for $\vec{q}$ has been confined to the first Brillouin zone.

When the temperature is very low, the interaction between two magnons can be neglected, and at the same time the self-energy effect disappears. However, as the temperature grows, the interaction between two magnons can not be ignored, and then the self-energy has to be taken into account. In fact, the renormalized magnon band can be experimentally detected via some well-established techniques, such as the inelastic neutron scattering, the inelastic X-ray scattering, the Brillouin light scattering and so on [42-47].

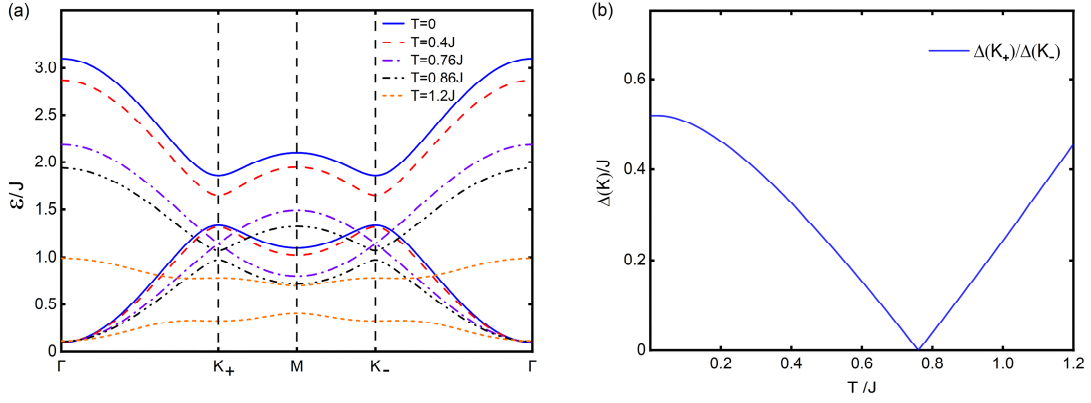

**Fig. 3.** The renormalized magnon band structures in the absence of the antichiral DMI: (a) The renormalized magnon dispersion curves along the path $\Gamma - K_+ - M - K_- - \Gamma$, (b) The gap $\Delta$ at the Dirac point $K_+$ and $K_-$ as a function of temperature. The parameters are chosen as $D = 0.1J$, $D' = 0$, $B = 0.1J$, and $S = \frac{1}{2}$.

In Fig. 3, we show the renormalized magnon band structures for the system in the absence of the antichiral DMI. It is clearly seen that, with the increase of the temperature, the renormalized magnon band gap at the Dirac points $K_+$ and $K_-$ decreases and will close at approximately $T = 0.76J$. If one further increases the temperature, then the renormalized magnon band gap reopens at $K_+$ and $K_-$ points and its width increases with $T$. Obviously, the widths of the band gaps at the Dirac points $K_+$ and $K_-$ are equal because of inversion symmetry. It is not difficult to display specifically that both the unperturbed and perturbed Hamiltonian have the inversion symmetry via checking the relationship $\sigma_x H(-\vec{k}) \sigma_x = H(\vec{k})$ with $H = H_0(\vec{k})$ or $H = H_1(\vec{k})$ [36].

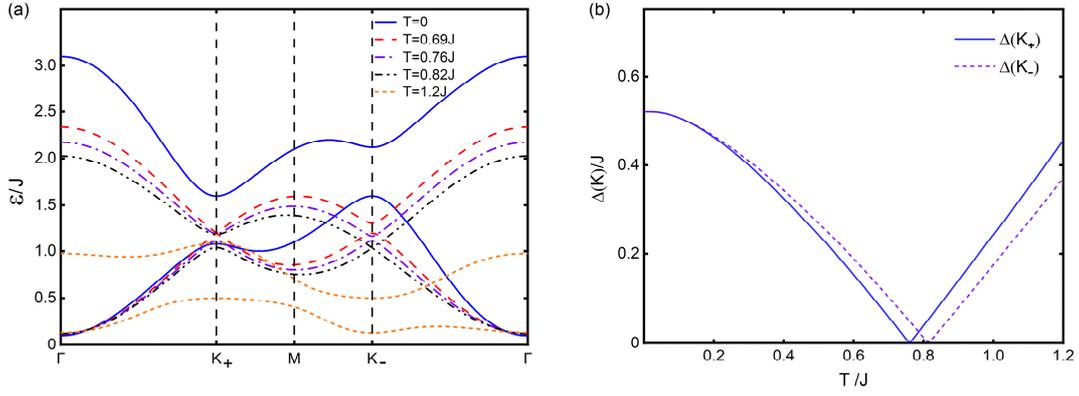

**Fig. 4.** The renormalized magnon band structures in the existence of the antichiral DMI: (a) The renormalized magnon dispersion curves along the path $\Gamma - K_+ - M - K_- - \Gamma$, (b) The gap $\Delta$ at the Dirac point $K_+$ and $K_-$ as a function of temperature. The parameters are chosen as $D = 0.1J$, $D' = 0.1J$, $B = 0.1J$, and $S = \dfrac{1}{2}$.

In Fig. 4, we display the renormalized magnon band structures for the system in the existence of the antichiral DMI. After introducing the antichiral DMI, the most interesting effect is that the band gaps at two non-equivalent Dirac points ($K_+$ and $K_-$) are no longer equal as temperature increases. It is in sharp contrast to the above result on the honeycomb ferromagnet without the antichiral DMI, where the band gaps are equal at two Dirac Dirac points [29,36,40,48]. As the temperature grows, the band gaps at the Dirac points $K_+$ and $K_-$ will close at $T_{K_+,c} \approx 0.69J$ and $T_{K_-,c} \approx 0.82J$, respectively.

*3.2. Topological properties of interacting Dirac magnon*

In bosonic systems, the occurrence of the nontrivial band topology is caused by the energy band structure. This nontrivial band topology can be characterized via a nonzero Berry curvature, which produces a quantized integer, i.e., the Chern number. Physically, a nontrivial band topology can arise only when the present system reveals a nontrivial gap between two magnon energy bands and each band possesses the nonzero Chern number [29]. In 2D bosonic systems, the Berry curvature can be expressed as

$$\Omega_\alpha(\vec{k}) = i\nabla_{\vec{k}} \times \langle \psi_\alpha(\vec{k}) | \nabla_{\vec{k}} | \psi_\alpha(\vec{k}) \rangle \qquad (10)$$

with $\alpha = u, d$. Here, $\psi_\alpha(\vec{k})$ and $\Omega_\alpha(\vec{k})$ represent the eigenvector and Berry curvature of the lower or the upper magnon band, respectively. By introducing a pseudospin freedom $\vec{\sigma}$, the Hamiltonian in Eq. (8) can be rewritten as

$$H_1(\vec{k}) = h_0 \sigma_0 + \vec{h} \cdot \vec{\sigma} \qquad (11)$$

with

$$h_0 = 3JS + B + 2D'S\rho_{\vec{k}} + Q_{\vec{k}} \qquad (12)$$

$$\vec{h} = \left(-\text{Re}\left((JS-M)\gamma_{\vec{k}} + p_{\vec{k}}\right), \text{Im}\left((JS-M)\gamma_{\vec{k}} + p_{\vec{k}}\right), m_{\vec{k}}\right) \qquad (13)$$

In the present representation, one can put the magnon band physics on one Bloch sphere, and all the geometric and topological natures are included in the total mass term $m_{\vec{k}} = 2DS\rho_{\vec{k}} - \delta_m$, which contains one scalar potential produced by the antichiral DMI. Physically, the scalar potential part can cause that the Berry curvatures of the Dirac points $K_+$ and $K_-$ are not equal. For the convenience of numerical calculation, one can adopt the Berry curvature of pseudospin freedom [49], which is

$$\Omega_\alpha(\vec{k}) = \mp \frac{1}{2h^3} \vec{h} \cdot \left(\frac{\partial \vec{h}}{\partial k_x} \times \frac{\partial \vec{h}}{\partial k_y}\right). \qquad (14)$$

where, $\vec{h} = (h_x, h_y, h_z)$, and $h = \sqrt{h_x^2 + h_y^2 + h_z^2}$.

The topological phase of the present system is characterized via the Chern numbers of each renormalized magnon band, which can be calculated by integrating the relevant Berry curvature over the whole first Brillouin zone (BZ), namely,

$$C_\alpha = \int_{BZ} d^2k\, \Omega_\alpha(\vec{k}). \tag{15}$$

Through adjusting the magnon population, we shall realize the topological phase transition in the upper magnon band from $C = -1$ to $C = 1$.

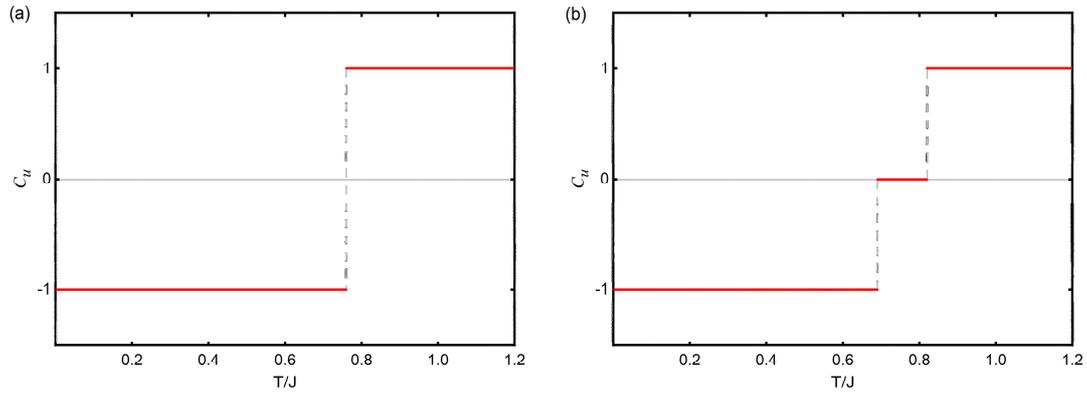

**Fig.5.** The Chern number for the upper band of the renormalized magnon: (a) the absence of the antichiral DMI ($D' = 0$), (b) the presence of the antichiral DMI ($D' = 0.1J$). The other parameters are chosen as $D = 0.1J$, $B = 0.1J$, and $S = \frac{1}{2}$.

In Fig. 5, we display the dependence of the Chern number for the upper band of the renormalized magnon on the temperature. Based on the previous results, we have now realized that the renormalized magnon gap can close and reopen at the critical temperature $T = T_c$, which means that there may exist the topological phase transition at $T_c$. From Fig. 5(a), in the absence of the antichiral DMI, one can clearly see the sign of that the Chern number for the upper band will change when $T > T_c$. In this case, the renormalized magnon gaps of the Dirac points $K_+$ and $K_-$ simultaneously close and reopen near the critical temperature $T_c$. By comparing the cases of $T < T_{K_+,c}$ with $T > T_{K_-,c}$, in the presence of the antichiral DMI, signs of the Chern number of the upper band in these two cases are different since the gap successively close and reopen at the Dirac points $K_+$ and $K_-$, as is shown in Fig. 5(b). When $T_{K_+,c} < T < T_{K_-,c}$, signs of Berry curvatures $\Omega_u(\vec{K}_+)$ and $\Omega_u(\vec{K}_-)$ are opposite. Thus, it is not hard to understand $C_u = 0$ when the temperature T is at the range $[T_{K_+,c}, T_{K_-,c}]$. Although there exists a topological gap at the Dirac point, in this case, the renormalized magnon band is topological trivial.

For magnons in the two-dimensional ferromagnet, one longitudinal temperature gradient can give rise to a transverse heat current via the Berry curvature [5,28]. Physically, such formation of the transverse heat current is referred to as the thermal Hall effect. The thermal Hall conductivity can be expressed as [5,28,50,51]

$$\kappa_{xy} = -\frac{k_B^2 T}{(2\pi)^2 \hbar} \sum_\alpha \int d^2k\, c_2(n_\alpha) \Omega_\alpha(\vec{k}). \tag{16}$$

Here, $n_\alpha = f(\varepsilon_\alpha(\vec{k})) = \left[e^{\beta \varepsilon_\alpha(\vec{k})} - 1\right]^{-1}$ corresponds to the famous Bose-Einstein distribution function, $\beta = 1/k_B T$, $c_2(x) = (1+x)\left(\ln\frac{1+x}{x}\right)^2 - (\ln x)^2 - 2\text{Li}_2(-x)$, and $\text{Li}_2(x)$ represents the dilogarithm function.

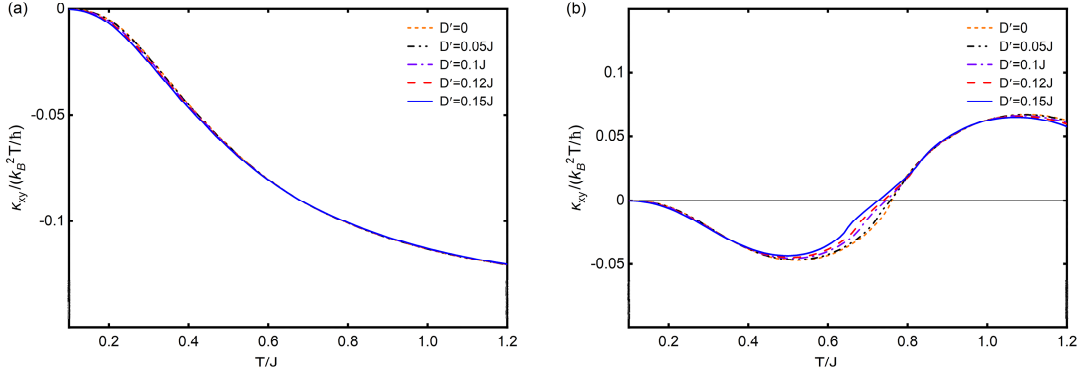

**Fig.8.** The thermal Hall conductivity vs temperature for different values of $D'$: (a) Neglecting magnon-magnon interactions, (b) Considering magnon-magnon interactions. The other parameters are set to $D = 0.1J$, $B = 0.1J$, and $S = \frac{1}{2}$.

In the linear spin wave approximation, i.e, neglecting magnon-magnon interactions, $\kappa_{xy}$ is always greater than zero, as displayed in Fig. 8(a). We note that the dependence of $\kappa_{xy}$ on the temperature $T$ is not significantly affected by varying the value of $D'$. In Fig. 8(b), one can clearly see that $\kappa_{xy}$ changes continuously as the temperature increases and goes from being negative at the low temperature $T$ ($T < T_c$) to being positive at the high temperature $T$ ($T > T_c$). In the absence of the antichiral DMI, the thermal hall conductivity and the Chern number of the upper band have the same signs with the increase of temperature. Most interestingly, their signs may be inconsistent when the antichiral DMI is introduced. According to our calculations, we infer that the critical temperature $T_c$ belongs to the range $[T_{K_+,c}, T_{K_-,c}]$ in the case of $D' \neq 0$. What is more, we find that the critical temperature $T_c$ goes down with increasing the value of $D'$. Physically, the sign reversal of the thermal Hall conductivity is an important indicator on one topological phase transition.

## 4. Conclusions

To conclude, we have presented a theoretical work on the topological phase transition in the Heisenberg ferromagnet possessing additional second nearest-neighbor DMIs on one honeycomb lattice. This unequal DMI between atoms on diverse sublattices can break the chiral symmetry of the honeycomb ferromagnet, which leads to a tilting of the magnon bands near the Dirac momenta. We have showed that the present ferromagnet can reveal different topological phases when the effect of magnon-magnon interaction is considered. It has been found that the topological phase of the present honeycomb ferromagnet can be controlled by changing the temperature. When the antichiral DMI is considered, the renormalized magnon band gaps at two non-equivalent Dirac points are no longer equal as temperature increases. Physically, the topological phase transition caused by magnon-magnon interactions are always accompanied with the occurrence of a band gap-closing. In the presence of the antichiral DMI, the critical temperature $T_c$ belongs to the range $\left[T_{K_+,c}, T_{K_-,c}\right]$. Our results have showed that the critical temperature $T_c$ goes down with increasing the value of $D'$.

**Acknowledgments**

We want to thank Dr. Hao Sun for many useful suggestions. This work was supported by the National Natural Science Foundation of China under Grant No. 12064011, the Natural Science Fund Project of Hunan Province under Grant No. 2020JJ4498 and the Graduate Research Innovation Foundation of Jishou University under Grant No. Jdy21030.